\documentclass[%
 aip,
 jmp,%
 amsmath,amssymb,
manuscript,%
]{revtex4-1}

\usepackage{graphicx}
\usepackage{epstopdf}
\usepackage{dcolumn}
\usepackage{bm}
\usepackage{booktabs}

\begin{document}


\title{Lower limits of line resistance in nanocrystalline Back End of Line Cu interconnects}

\author{Ganesh Hegde}
	\email{ganesh.h@ssi.samsung.com}
\author{R. Chris Bowen}%
\author{Mark S. Rodder}
\affiliation{Advanced Logic Lab, Samsung Semiconductor Inc., Austin, TX 78754, USA}
\date{\today}
\begin{abstract}
The strong non-linear increase in Cu interconnect line resistance with decreasing linewidth presents a significant obstacle to their continued downscaling. In this letter we use first principles density functional theory based electronic structure of Cu interconnects to find the lower limits of their line resistance for metal linewidths corresponding to future technology nodes. We find that even in the absence of scattering due to grain boundaries, edge roughness or interfaces, quantum confinement causes a severe increase in the line resistance of Cu. We also find that when the simplest scattering mechanism in the grain boundary scattering dominated limit is added to otherwise coherent electronic transmission in monocrystalline nanowires, the lower limit of line resistance is significantly higher than projected roadmap requirements in the International Technology Roadmap for Semiconductors.
\end{abstract}

\maketitle
Damascene deposited Copper has been the Back End of Line (BEOL) interconnect conductor of choice for semiconductor technology nodes since 2000. Copper has several desirable features, including a low bulk resistivity ($\approx$ 1.68 $\mu \Omega$ cm at room temperature, second only to Ag among elemental metals \cite{giancoli2008physics}) and the ability to easily be electro-deposited and annealed at temperatures compatible with semiconductor processing. 

With metal linewidth scaling proceeding in sync with technology scaling, it should be expected that the resistance per unit length (henceforth referred to as $R/l$ with units in $\Omega/\mu$m) of BEOL Cu metallization will increase. It has been known for some time, however, that metal resistivity also increases with a decrease in metal dimensionality \cite{sondheimer1952mean}. This adds to the increase in $R/l$ that is expected from a simple reduction in dimensions. This increase in metal resistivity has been attributed to an increase in electron scattering due to grain boundaries \cite{mayadas1970electrical}, surfaces and interfaces \cite{fuchs1938conductivity,sun2009dominant, sun2010surface}. Consequently, the International Technology Roadmap for Semiconductors (ITRS) has projected an increase in roadmap effective metal resistivity (for the lowest BEOL metallization level or M1) from 4.77 $\mu \Omega$-cm in 2016 to 11.41 $\mu \Omega$-cm by the year 2028 \cite{wilson2013international}. This corresponds to a significant increase in projected $R/l$ requirement from approximately 35 $\Omega/\mu$m in the year 2016 to 1180 $\Omega/\mu$m in the year 2028.

To determine if Cu can meet this requirement, we use first principles Density Functional Theory (DFT) based electronic structure simulations of Cu to find the lower limits of its line resistance $R/l$ for metal linewidths corresponding to future technology nodes. These lower limits are then compared to ITRS projected requirements for the lowest Back End of Line metallization level (M1)\cite{wilson2013international}.

The motivation for a lower limits investigation is that if the $R/l$ lower limits are sufficiently below projected requirements, then they can be achieved through continued improvements in BEOL resistance engineering. If the lower limits themselves are significantly above projected requirements, it serves to re-evaluate if Cu is suitable for future nodes. Our investigation is similar in spirit to previous calculations on the lower limits of contact resistivity in Si \cite{maassen2013full, hegde2014effect}. 

We wish to state explicitly what we mean by 'lower limits of line resistance' in this letter. In the absence of any electron scattering mechanism such as that due to phonons, impurities, interfaces and grain boundaries, conductance is decided simply by the number of conducting channels available. If we translate this definition for the case of Cu interconnects, we infer that the ballistic/Landauer limit of conductance of monocrystalline nanowires in the coherent transport regime represents the upper limit of Cu interconnect conductance and their corresponding resistance represents the lower limit \cite{datta2005quantum}. While the resistance of monocrystalline nanowires represents the theoretical lower limit, they do not (yet) represent a practically achievable lower limit. The processes involved in the formation of Cu interconnects result in, at best, structures with a high concentration of low specific resistivity [111]/[111] $\Sigma 3$ twin boundaries of Coincidence Site Lattice (CSL) type \cite{lu2004ultrahigh}. A practically achievable lower limit would therefore consist of nanowires with such twin boundaries along the direction of transport.

In this letter we therefore report both theoretical and practical lower $R/l$ limits. The rest of this letter is organized as follows. We first present the details of our computational method, followed by line resistance calculations versus cross sectional area for monocrystalline nanowires oriented along 6 different transport directions. Next, the simplest grain boundary scattering mechanism - twin [111]/[111] $\Sigma$3 boundaries are added to [111] oriented nanowires and their $R/l$ versus cross sectional area is compared to ITRS requirements. Finally, we compare the line resistance between these ideal twin structures and their more realistic nanocrystalline counterparts. We then conclude with a discussion on the implications of our lower-limits study.  

We note here that DFT has been shown in past literature to accurately reproduce the bulk Fermi Surface and resistivity of metal nanostructures\cite{feldman2010simulation, CuModel, CuConfinement, kim2010large, ke2009resistivity, timoshevskii2008influence, zhou2010ab}. The usual procedure followed in several of the aforementioned studies is the computation of the self-consistent electronic density of the metal structures, followed by a Landauer-Buttiker analysis of electrical conductance \cite{datta2005quantum}. This is the procedure we follow in this work as well. We first compute ballistic conductance of monocrystalline Cu nanowire structures oriented along six transport orientations - [100], [110], [111], [112], [120] and [122] - using the Landauer-Buttiker formalism. The cross sectional areas of the monocrystalline structures studied ranged from 20 to 100 nm$^2$. The choice of orientations was motivated by the dominant orientations of grains seen in the direction of transport in damascene deposited Cu interconnects \cite{ganesh2012effect}. The ballistic conductance in this formalism is given as G = $\frac{2e^2}{h}T$ where
\begin{equation}
T = \int T(E)\frac{-\partial f}{\partial E}dE
\end{equation}
where $E$ is energy, $\frac{2e^2}{h}$ is the quantum of conductance, $T(E)$ is the net number of modes at energy $E$ and $f$ is the Fermi function). The net number of modes $T$ is obtained by computing the self-consistent band structure and simply counting the number of conducting modes at each energy step broadened by the Fermi function.
Cu nanowires with cross sectional areas of 100 nm$^2$ contain between 5000 to 15000 atoms (depending on the orientation). Computing the self consistent band structure of such a large number of atoms in DFT is computationally cumbersome. To enable scaling of DFT calculations to the required system size, a recently published $ab$ $initio$ s-orbital approximation to the basis of Cu was used to compute the self consistent band structure of the nanowires \cite{hegde2015feasibility}. For additional details of the computational procedure, we refer the interested reader to this work.
Ballistic conductance is independent of the length of the cross section in the transport direction. To compute line resistance per unit length $R/l$ a length scale corresponding to a scattering event is introduced. Here we assume a scattering length equal to bulk inelastic mean free path in Cu (400\AA). It has been shown \cite{datta2005quantum} that an introduction of a scattering mean free path in the Landauer-Buttiker formalism modifies the ballistic conductance G as follows
\begin{equation} 
G = \frac{2e^2}{h}T\frac{\lambda}{\lambda+l}
\end{equation}
where $\lambda$ is the scattering mean free path, which we approximate to be independent of energy. Since M1 interconnect run lengths ($l$) range in microns, we have $l >> \lambda$, this results in
\begin{equation}
\frac{R}{l} = \frac{h}{2e^2}\frac{1}{T\lambda}
\end{equation}

Using the bulk mean free path is a rather generous assumption, given that the width of the nanowires investigated is a maximum of 10 nm. It is thus safe to say that the $R/l$ numbers for monocrystalline Cu nanowires represent the lower limit of line resistance in Cu interconnects.
 
To compare our simulations to ITRS projections, we used projected Cu requirements for resistivity data available in the 2013 ITRS table (see MPU Interconnect Technology requirements for M1 level in the roadmap \cite{wilson2013international}) and calculate projected $R/l$ requirements for cross sectional areas between 20 to 100 nm$^2$. Damascene Cu lines are bounded by a cladding/barrier layer on two sidewalls and the trench bottom. We subtracted the barrier/cladding thickness outlined in the ITRS table from the trench bottom and from each sidewall to obtain the effective Cu cross sectional area $A$. The $R/l$ ITRS projection was then computed by dividing the conductor effective resistivity $\rho$ (obtained from aforementioned table) by $A$ in line with Ohm's law.

\begin{figure}
	\centering
		\includegraphics[width=3in, height=2.5in]{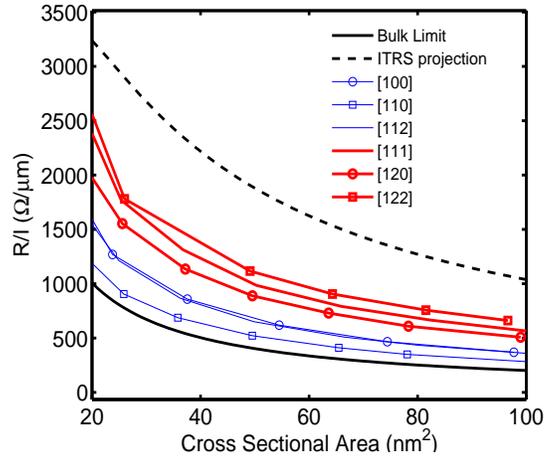}
	\caption{Comparison of $R/l$ lower limits for monocrystalline Cu nanowires with different orientations to ITRS requirements. Bulk Cu $R/l$ computed with the Fermi surface averaged transmission is also shown for reference.}
	\label{fig:R_lower_limits_paper}
\end{figure}

\begin{figure*}[!htbp]
	\centering
	\includegraphics[width=\textwidth,height=2.0in]{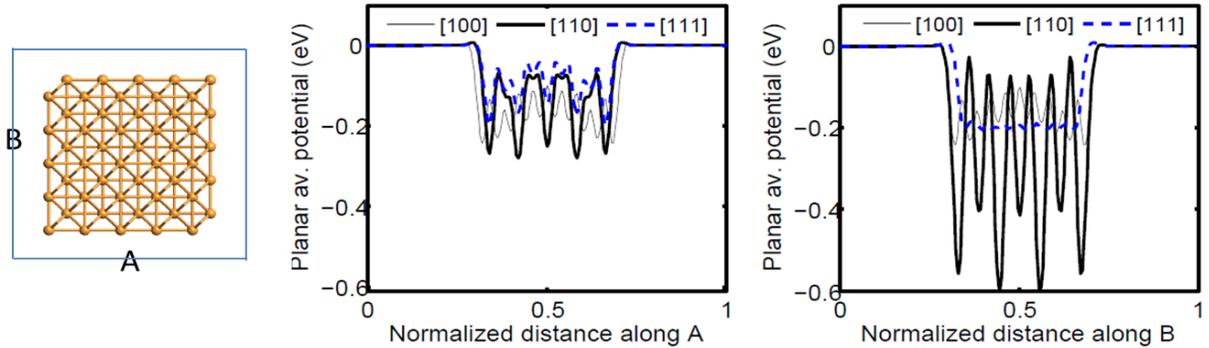}
	\caption{Planar averaged electrostatic potential profile for nanowires having a cross sectional area of 1 nm$^2$ oriented along [100], [110] and [111]. A and B represent the orthogonal directions of confinement and are mutually orthogonal to the direction of transport indicated in the figure legend.}
	\label{fig:Potential_Profiles}
\end{figure*}

Figure \ref{fig:R_lower_limits_paper} shows a comparison of $R/l$ lower limits for different Cu orientations obtained from DFT and $R/l$ ITRS projected requirements for Cu cross sectional areas (i.e. without liner and barrier) between 20 and 100 nm$^2$. For the sake of reference, the bulk $R/l$ lower limit (obtained by averaging the bulk ballistic transmission over the Fermi surface of Cu) is also plotted. It is evident that the lowest $R/l$ limit in Cu is determined by [110] oriented monocrystalline nanowires. It can also be seen that this lowest $R/l$ limit in monocrystalline Cu [110] is well below ITRS projected requirements for all cross sectional areas shown.

The mismatch in line resistance between nanowire orientations comes from the anisotropy in the ballistic conductance between orientations that has been discussed in previous publications \cite{CuConfinement,jones2015electron}. The conductance anisotropy under confinement for a prototypical metal such as Cu that has an approximately spherical bulk Fermi surface can in turn be explained by reference to the planar averaged electrostatic potential in the nanowires along the direction of confinement. Figure \ref{fig:Potential_Profiles} shows such a comparison for three nanowire orientations of equal cross sectional area. It is evident that the confining potential varies quite strongly for the nanowires oriented differently. This anisotropy in quantum confinement potential leads to an anisotropy in conducting modes. 

As mentioned previously, while the calculation of $R/l$ for monocrystalline nanowires may indicate a theoretical limit that is lower than ITRS projected requirements, we are unaware of a practical method to pattern perfectly monocrystalline metal interconnects on a large scale required for semiconductor technology. Existing techniques such as electrodeposition used in the damascene process result in interconnects that are polycrystalline with a wide variation in the grain orientation distribution \cite{ganesh2012effect}. It has, however, been demonstrated that Cu nanowires with a high degree of 'twinning' along the [111] transport direction can be fabricated \cite{lu2004ultrahigh}. Recent DFT calculations of the specific resistivity of grain boundaries also indicate that the [111]/[111] $\Sigma$3 Coincidence Site Lattice (CSL) or twin boundary has the least specific resistivity among all grain boundaries \cite{cesar2014calculated}. This has been attributed to negligible bond-orientational disorder seen at such boundaries compared to other grain boundaries. It is therefore reasonable to infer that the $R/l$ values of nanowires with [111]/[111] $\Sigma$3 twin boundaries along transport direction represent a more practically achievable lower limit.

\begin{figure}
	\centering
		\includegraphics{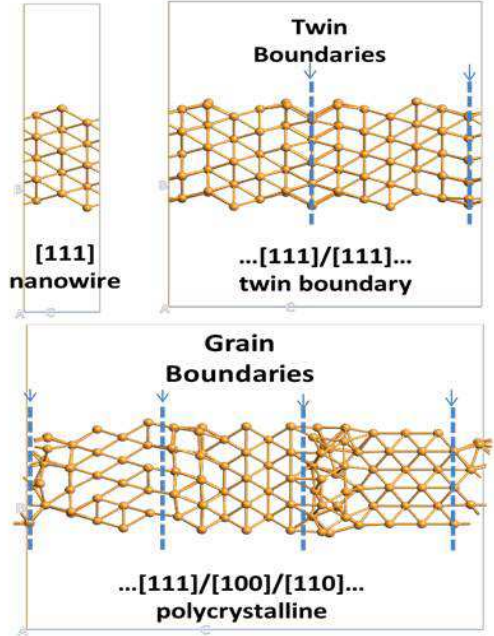}
	\caption{Examples of three categories of structures simulated in this work. From left to right, these are monocrystalline nanowires, twin boundary $\Sigma 3$ type nanowires and polycrystalline nanowires with grain size equal to linewidth and varying grain orientation distribution.}
	\label{fig:PaperStructures}
\end{figure}

The calculations performed on monocrystalline nanowires were therefore repeated for nanowires consisting of [111]/[111] twin boundaries along the direction of transport. Since a significant impact of Aspect Ratio (AR) was not seen for monocrystalline nanowires, we limited the calculations to an AR of 1$\times$1. Figure \ref{fig:PaperStructures} shows an example of such a twin boundary structure. In keeping with data from recent experiments and other simulation studies of the Cu interconnect system \cite{roberts2015resistivity, pyzyna2015resistivity}, grain size was assumed to equal line width. As shown in figure \ref{fig:PaperStructures}, the eventual structure simulated consists of twin grains infinitely repeated along the direction of transport. $R/l$ values for cross sectional areas up to 60 nm$^2$ (system size of up to 25000 atoms) were fit and projected up to cross sectional areas of 100 nm$^2$. 

\begin{figure}
	\centering
		\includegraphics[width=3in,height=2.5in]{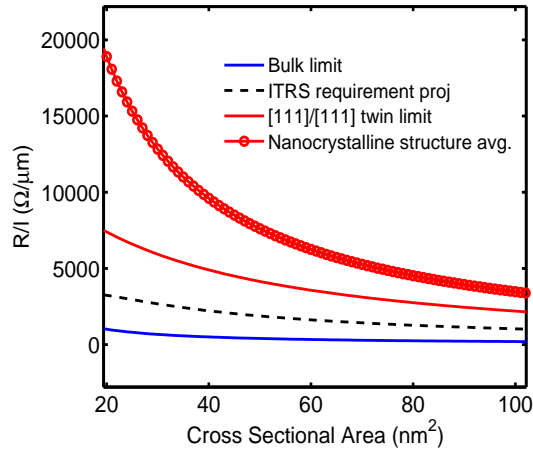}
	\caption{A comparison of the $R/l$ values for twin boundary structures, polycrystalline structures and ITRS requirements. Also included in the comparison is the $R/l$ for bulk Cu.}
	\label{fig:R_by_l_111_poly_limits}
\end{figure}

The results of the $R/l$ calculation for twin boundary structures and its comparison to ITRS projections is shown in figure \ref{fig:R_by_l_111_poly_limits}. It is very clear that a change from [111] oriented monocrystalline to the simplest boundary - that of [111]/[111] twin structures - results in an increase in $R/l$ values to the extent that these are now higher than projected ITRS requirements. This is even after discounting the fact that we used the bulk scattering mean free path for the twin boundary structures even though the grain size was significantly smaller than the mean free path used.

It must be emphasized that the twin boundary structures considered in this study are highly ordered. Process related non-idealities such as grain orientation anisotropy (GOA), line edge roughness, and the presence of a Cu-liner interface will result in actual $R/l$ values that are higher than this ideal case. GOA, especially, will result in a significant resistance penalty owing to increased anisotropy in conductance for different orientations when Cu is quantum confined as discussed previously. To estimate the impact of more realistic structures on line resistance, we simulated a large number of polycrystalline structures with a wide variety of grain orientation distributions for each cross sectional area. We retained the structural assumptions regarding aspect ratio and grain size made for twin boundaries. An example of such a structure is shown in figure \ref{fig:PaperStructures}. We then averaged the ballistic conductance for all of these structures and computed an averaged $R/l$ value representative of all structures simulated. The resultant line resistance shown in figure \ref{fig:R_by_l_111_poly_limits} is significantly higher than the $R/l\propto$ limits of twin boundary structures and ITRS projections.

The results presented in this letter indicate that from a materials perspective, nanocrystalline Cu will not be able to meet projected ITRS requirements.  While remedial measures such as reduction in GOA, increase in grain size and a reduction in barrier thickness may improve Cu line resistance significantly, it is not possible to completely eliminate either grain boundaries or barriers and liners in the current damascene Cu BEOL paradigm. Additionally, if BEOL scaling is to be maintained, then confinement and its concomitant effects such as the severe rise in line resistance shown previously are unavoidable. Even the best case that can practically be engineered - that of $\Sigma$3 twin boundaries - does not seem to meet projected requirements. While a exhaustive exploration of all possible grain orientation distributions is not feasible even in simulation, the $R/l$ values of the polycrystalline structures we calculated serve to illustrate the extent to which the aforementioned structural imperfections increase line resistance well beyond ITRS requirements.

In conclusion, we reported the results of our investigation into the lower limits of line resistance in Cu interconnects. We found that simple confinement results in a significant increase in Cu line resistance. When the simplest grain boundary scattering mechanism - scattering at highly ordered [111]/[111] $\Sigma$3 twin boundaries - was introduced into otherwise ballistic monocrystalline [111] interconnects, we found that the $R/l$ limits increase beyond ITRS projected requirements. This increase in line resistance presents a significant obstacle to the continued use of Cu interconnects in future technology nodes. We believe that the results of this investigation warrant increased attention on schemes and materials that mitigate the potentially rapid rise in BEOL resistance in future technology nodes.    

We thank Titash Rakshit, Borna Obradovic and Rwik Sengupta for helpful suggestions made during the course of this work.
\nocite{*}
%

\end{document}